\documentclass[journal]{IEEEtran}
\usepackage[dvipdfmx]{graphicx,xcolor}
\usepackage{epsfig}
\usepackage[T1]{fontenc}
\usepackage{lmodern}
\usepackage{textcomp}
\usepackage{latexsym}
\usepackage{cite}
\usepackage{bm}
\usepackage{amsmath}
\usepackage{amsfonts}

\newcommand{\vct}[1]{\mbox{\boldmath{$#1$}}}

\newcommand{\ee}        {\mathrm{e}}
\newcommand{\jj}        {\mathrm{j}}
\newcommand{\dd}        {\mathrm{d}}
\newcommand{\TT}        {\mathrm{T}}
\newcommand{\HH}        {\mathrm{H}}

\ifCLASSINFOpdf
\else
\fi
\begin{document}
\title{Radar-Based Respiratory Measurement of \\ a Rhesus Monkey by Suppressing Nonperiodic \\ Body Motion Components}

\author{Takuya~Sakamoto,~\IEEEmembership{Senior Member, IEEE,} Daisuke~Sanematsu,
  Itsuki~Iwata,~\IEEEmembership{Student Member, IEEE,} Toshiki~Minami, and~Masako~Myowa
\thanks{T.~Sakamoto, D.~Sanematsu, and I.~Iwata are with the Department of Electrical Engineering, Graduate School of Engineering, Kyoto University, Kyoto 615-8510, Japan.}
\thanks{T.~Minami and M.~Myowa are with Graduate School of Education, Kyoto University, Kyoto 606-8501, Japan.}}
\markboth{}%
         {Sakamoto \emph{et al.}: Radar-Based Respiratory Measurement of a Rhesus Monkey by Suppressing Nonperiodic Body Motion Components}


\maketitle  
  \begin{abstract}
    We propose a method to measure the respiration of a rhesus monkey using a millimeter-wave radar system with an antenna array. Unlike humans, small animals are generally restless and hyperactive in nature, and suppression of their body motion components is thus necessary to realize accurate respiratory measurements. The proposed method detects and suppresses nonperiodic body motion components while also combining and emphasizing the periodic components from multiple echoes acquired from the target. Results indicate that the proposed method can measure respiration rate of the target monkey accurately, even with frequent body movements.
\end{abstract}

\IEEEpeerreviewmaketitle


\section{Introduction}
\IEEEPARstart{R}{espiratory} patterns in both humans and animals are known to be affected by mental stress and health conditions. Measurement of an animal's respiration can play an important role in detecting early signs of mental stress, respiratory infections, and other health conditions. Although contact-type respiratory sensors are commonly used for medical purposes, these sensors are not ideal for animal monitoring because of the discomfort caused by wearing them.

Recently, noncontact respiratory sensing using radar systems has been studied intensively. The first report of respiratory measurement of a rabbit using a radar system was published as early as 1975 \cite{JC_Lin1975} and was followed by various studies for respiratory measurement of animals including a hibernating black bear \cite{S_Suzuki2009}, a horse \cite{T_Matsumoto2022}, and a cow \cite{SA_Tuan2022}. These animals are relatively gentle and do not make frequent body movements, which means that use of conventional techniques \cite{M_Nosrati2019,SMM_Islam2020,X_Shang2020,J_Xiong2020,AA_Pramudita2021,C_Feng2021,M_Mercuri2021,Harikesh2021,T_Zheng2022} for radar-based respiratory measurements is feasible. However, it remains challenging to measure smaller animals that make frequent body movements because these movements interfere with the small body displacements caused by respiration, thus degrading the measurement accuracy.

To achieve accurate noncontact physiological measurements with the presence of these body movements, various techniques and systems have been proposed \cite{A_Singh2021}. These approaches include a DC offset calibration method \cite{C_Li2008}, a self-injection-locked architecture \cite{SH_Yu2020}, a multi-channel Kalman smoother \cite{Q_Wu2021}, a frequency-locked loop radar system \cite{KC_Peng2021}, and a body movement cancelation technique using two radar systems on the front and rear sides of the target body \cite{J_Zhang}. Despite these efforts, the effects of body movements are still problematic, particularly when measuring a small animal such as a rhesus monkey that makes frequent movements.

In this study, we propose a new method to suppress body motion components using short-time autocorrelation functions of the displacement waveforms, and apply this method to radar echo signals acquired from a rhesus monkey that makes frequent body movements. To improve respiratory measurement accuracy, the proposed method combines multiple respiratory intervals estimated at multiple positions on the radar image to emphasize the periodic components related to respiration. The experimental results show that the respiratory intervals estimated using a pair of radar systems were in good agreement when the proposed method was used, indicating the effectiveness of our technique.

\section{Proposed Method for Respiratory Measurement}
We use two sets of frequency-modulated continuous-wave (FMCW) radar systems in combination with a multiple-input multiple-output (MIMO) antenna array. The MIMO antenna array, which contains $K_\mathrm{T}$ transmitting and $K_\mathrm{R}$ receiving elements, can be approximated using a linear array consisting of $K=K_\mathrm{T}K_\mathrm{R}$ virtual elements, as long as there are no overlapping virtual elements.

Assuming that the linear antenna array contains $K$ virtual elements, $s_k(t,r)$ denotes the signal received by the $k$th virtual element $(\mathrm{where}\; k=0,1,\cdots,K-1)$, where $t$ is the slow time, and the range $r$ is expressed as $r=ct'/2$ using the fast time $t'$ and the speed of light $c$. A signal vector $\vct{s}(t,r)$ is defined as $\vct{s}(t,r) = [s_0(t,r), s_1(t,r), \cdots , s_{K-1}(t,r)]^\TT$, where the superscripted T represents a transpose operator. Note that because an FMCW radar system is used here, the fast time $t'$ can be obtained by simply converting the beat frequency that is generated by mixing the received signal with the transmitted signal.

Let us define two-dimensional Cartesian coordinates $(x,y)$, where the array baseline is located on the $x$-axis and the $x$ coordinate of the $k$th element of the array is assumed to be $x_k$. We also assume that all targets are located within the half-plane $y\geq 0$.

Using the beamformer weight $w_k(\theta)=\alpha_k\ee^{\jj (2\pi x_k/\lambda)\cos\theta}$ $(k=0,1,\cdots,K-1)$ for the angle $\theta$, the weight vector can be defined as $\vct{w}(\theta)=[w_0,w_1,\cdots,w_{K-1}]^\mathrm{T}$. Here, $\lambda$ is the wavelength and $\alpha_k$ is a Taylor window coefficient.

We generate a complex radar image $I'(t,\vct{r})=\vct{w}(\theta)^\HH \vct{s}(t,r)$, where the superscripted H represents a conjugate transpose operator and $\vct{r}$ is the position vector, which can be expressed as $(r,\theta)$ in polar coordinates. Because the complex radar image $I'(t,\vct{r})$ contains static clutter, the time-averaged component is subtracted as
$I(t,\vct{r})=I'(t,\vct{r})-({1}/{T})\int_{0}^{T} I'(t,\vct{r}) \,\dd t,$
where $T$ is the time for which the target is approximated to be almost stationary. Note that during actual implementation as digital data, $r$ and $\theta$ are discretized as $r_1, r_2, \cdots$ and $\theta_1, \theta_2, \cdots$, respectively, which results in realization of discretized position vectors $\vct{r}_1,\vct{r}_2,\cdots,\vct{r}_M$.

First, the intensity of the complex radar image is time-averaged as
$\bar{I}(\vct{r})=({1}/{T}) \int_{0}^{T} \left|I(t,\vct{r})\right|^2 \dd t,$
and $\vct{r}_0$, which is the target location, is estimated from $\bar{I}(\vct{r})$ as
$\vct{r}_0 = \arg\max_{\vct{r}}\bar{I}(\vct{r}).$
Then, a non-empty connected region $R_0\subseteq\mathbb{R}^2$ is determined such that it satisfies $\vct{r}_0\in R_0$ and also
\begin{equation}
  R_0 = \left\{\vct{r}\subseteq\mathbb{R}^2\left|\bar{I}(\vct{r})\geq \eta \right.\right\},
  \label{eq:4}
\end{equation}
which means that the radar image values are equal to or larger than a threshold $\eta$. The displacement $d_m(t)$ of the $m$th position vector ($m=1,2,\cdots,M$) is then estimated for $\forall \vct{r}_m\in R_0$ as $\hat{d}_m(t) = ({\lambda}/{4\pi}){\rm unwrap}\left(\angle I(t,\vct{r}_m)\right),$ where $\angle$ denotes the phase of the complex number and ${\rm unwrap}(\cdot)$ denotes a phase unwrapping operator. Then, we calculate a line-of-sight velocity $\hat{v}_m(t)$ as $\hat{v}_m(t) = (\dd/\dd t)\hat{d}_m(t)$. Similarly, the estimated velocity $\hat{v}_0(t)$ can be obtained from the estimated target location $\vct{r}_0$ as $\hat{v}_0(t) = {\lambda}/{4\pi}({\dd}/{\dd t}){\rm unwrap}\left(\angle I(t,\vct{r}_0)\right)$.

A short-time auto-correlation function composed of $\hat{v}_m(t)$ is defined as:
\begin{equation}
  \rho_m(t,\tau) = \frac{1}{\sqrt{D}}\int_{-T_0/2}^{T_0/2} \hat{v}_m(t'-t)\hat{v}_m(t'-t-\tau) \dd t',
  \label{eq:6}
\end{equation}
where $T_0$ is a window width and $D$ is obtained as follows: 
\begin{equation}
  D = {\int_{-T_0/2}^{T_0/2} \left|\hat{v}_m(t'-t)\right|^2 \dd t'\int_{-T_0/2}^{T_0/2} \left|\hat{v}_m(t''-t-\tau)\right|^2 \dd t''}.
    \label{eq:7}
\end{equation}

Using ~(\ref{eq:6}), we calculate the estimated respiratory interval $\hat{\tau}_m$ as $\hat{\tau}_m(t) = \arg\max_{\tau} h(\tau)\rho_m(t,\tau),$ where $h(\tau)$ is a Tukey window function that covers the time lag range $\tau_{\mathrm{S}}\leq\tau\leq\tau_{\mathrm{L}}$ with $\tau_{\mathrm{S}}=0.8$ s and $\tau_{\mathrm{L}}=2.0$ s. This range was determined based on the typical respiratory rate for the subject animal. Note that the estimated respiratory interval $\hat{\tau}_m$ is also dependent on $t$ because $\hat{v}_m(t)$ is quasiperiodic rather than periodic, and thus we write $\hat{\tau}_m(t)$ to indicate this dependency explicitly.

In many conventional methods, the respiratory interval is estimated from the signal for $\vct{r}_0$ that corresponds to the maximum peak of the radar image $\bar{I}(\vct{r})$. This method is called the conventional method hereafter for comparison purposes. As will be explained later, this method can easily be affected by the body movements of the target animal/person. This is why we propose a new method, which will be explained below.

If the velocity $\hat{v}_m(t)$ is a quasiperiodic function of $t$, then its short-time correlation function $\rho_m(t,\tau)$ is also quasiperiodic in the direction of the time lag $\tau$, with its largest peak $\rho_m(t,0)=1$ occurring at $\tau=0$ and its second largest peak occurring at $\tau=\tau_m$ ($\tau_m>0$), where $\tau_m$ is equal to the instantaneous respiratory interval if the displacement $\hat{v}_m(t)$ is due to periodic respiration alone.

In contrast, if $\hat{v}_m(t)$ is affected by body motion, then $\hat{v}_m(t)$ cannot be regarded as a quasiperiodic function and the second peak $\tau_m$ cannot be used as an estimate of the respiratory interval, which illustrates the importance of automatic detection of the local periodicity of $\hat{v}_m(t)$.

To detect the periodicity of $\hat{v}_m(t)$, we use the waveform $\rho_m(t,\tau)$ and fit $\rho_m(t,\tau)$ using a cosine function $\cos(2\pi \tau/\tau_m)$ as follows: 
\begin{equation}
  \varepsilon_m(t) =\min_{\tau_m}\frac{1}{\tau_0}\int_0^{\tau_0}\left| \rho_m(t,\tau)-\cos(2\pi \tau/\tau_m)\right|^2\ \dd \tau.
\end{equation}
We then use $\varepsilon_m(t)$ as an indicator of the periodicity; $\tau_m(t)$ can be trusted as an estimate if $\varepsilon_m(t)$ is small, whereas $\hat{v}_m(t)$ is likely to contain a body motion component if $\varepsilon_m(t)$ is large.

Our proposed method estimates the respiratory interval $\hat{\tau}(t)$ using
\begin{equation}
  \hat{\tau}(t) = \frac{\displaystyle\sum_{\vct{r}_m\in R_0} \frac{\hat{\tau}_m(t)}{\varepsilon_m(t)}}{\displaystyle\sum_{\vct{r}_m\in R_0} \frac{1}{\varepsilon_m(t)}},
\end{equation}
where the inverse number of $\varepsilon_m(t)$ is used as a weight in the weighted average of the respiratory interval estimates $\hat{\tau}_m(t)$ for multiple position vectors $\vct{r}_m\in R_0$. In addition, if the denominator is below a threshold $\varepsilon_{\mathrm{th}}$ such that
\begin{equation}
{\displaystyle\sum_{\vct{r}_m\in R_0} \frac{1}{\varepsilon_m(t)}} < \frac{M}{\varepsilon_{\mathrm{th}}},
\end{equation}  
the estimated respiratory interval $\hat{\tau}(t)$ is not used because the displacement waveform cannot be regarded as being quasiperiodic. The next section evaluates the proposed method's performance using measured radar data acquired from a hyperactive rhesus monkey that lives in a zoo.

\section{Radar Measurement Setup with Monkey}
We used a pair of millimeter-wave array radar systems in this setup. Both systems are FMCW radar systems with a center frequency of 79 GHz, a center wavelength of $\lambda=3.8$ mm, and a bandwidth of 3.6 GHz. The beamwidths of the transmitting elements are $\pm4^\circ$ and $\pm33^\circ$ in the E- and H-planes, respectively; the beamwidths of the receiving elements are $\pm4^\circ$ and $\pm45^\circ$ in the E- and H-planes, respectively.

The radar array is composed of a MIMO array that contains three transmitting and four receiving elements, with 7.6 mm ($2\lambda$) spacings between the transmitting elements and 1.9 mm ($\lambda/2$) spacings between the receiving elements. In the experimental setting, the MIMO array can be approximated using a virtual linear array. Specifically, the array in this study can be approximated using a 12-element virtual linear array with element spacings of $\lambda/2$. The slow-time sampling frequency was 100 Hz. The two radar systems were separated from each other by a distance of $0.7$ m, as described in the previous section.

The accuracy of respiratory interval estimation is evaluated by comparing estimates obtained from the two radar systems because it is difficult to attach a contact-type respiration sensor to animals such as monkeys. Note that when the accuracy is evaluated in this way, it may appear to be higher than the actual accuracy because the estimation errors for the two radar systems may be correlated. Despite this, the lower bound of the accuracy can be evaluated using our approach. Providing a more precise evaluation of the accuracy will be the next step in our future studies.

\begin{figure}[bt]
  \begin{center}
    \includegraphics[width=0.8\linewidth]{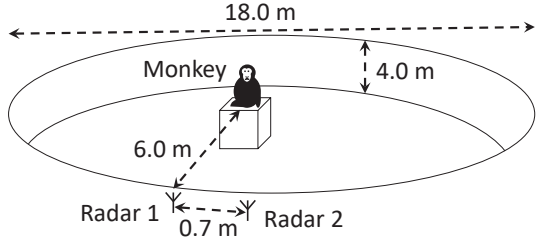}
    \caption{Schematic of the measurement setup with two radar systems and a monkey located in a cylindrical enclosure.}
      \label{fig1}
  \end{center}
\end{figure}

\begin{figure}[bt]
  \begin{center}
    \includegraphics[width=0.55\linewidth]{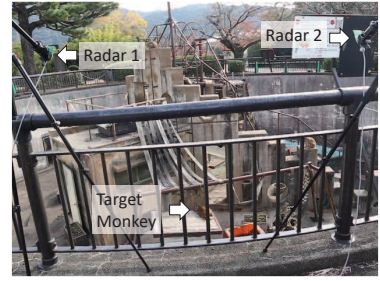}
    \caption{Photograph of the measurement environment with the two radar systems.}
      \label{fig2}
  \end{center}
\end{figure}

We measured the respiration of a rhesus monkey at the Kyoto City Zoo, where twelve rhesus monkeys were housed in a cylinder-shaped monkey enclosure with an artificial mountain made from a concrete pile with blocks of various shapes. The enclosure diameter is $18.0$ m and its depth is $4.0$ m. We performed radar measurements of a target monkey that was located away from the others to avoid interference caused by radar echoes from the other monkeys. The measurement time was $T=120$ s, and during this time period, the target monkey was located approximately 6 m away from the radar system. A schematic and a photograph of the measurement setup are shown in Figs. \ref{fig1} and \ref{fig2}, respectively.

\section{Experimental Performance Evaluation of the Proposed Method}
\begin{figure}[bt]
  \begin{center}
    \includegraphics[width=0.7\linewidth]{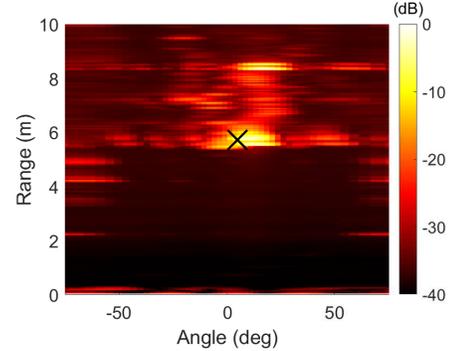}
    \caption{Radar image $\bar{I}(\vct{r})$ after static clutter suppression with a cross symbol indicating the estimated target location $\vct{r}_0$.}
      \label{fig4}
  \end{center}
\end{figure}

\begin{figure}[bt]
  \begin{center}
    \includegraphics[width=0.7\linewidth]{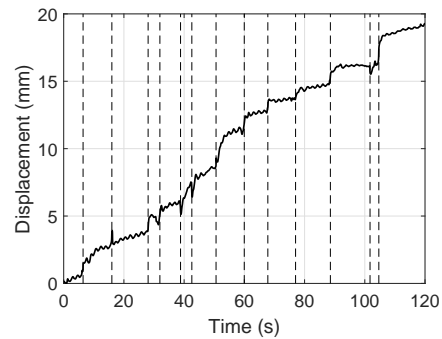}
    \caption{Estimated displacement $\hat{d}_0(t)$ versus time, where dashed lines indicate body movements.}
      \label{fig6}
  \end{center}
\end{figure}

Fig. \ref{fig4} shows a time-averaged radar image $\bar{I}(\vct{r})$ after suppression of static clutter, where the maximum peak is indicated by a cross-shaped symbol at a range of $5.72$ m and an angle of $9.6^\circ$; this peak corresponds to the estimated location of the target $\vct{r}_0=\arg\max_{\vct{r}}\bar{I}(\vct{r})$. Using the estimated target location $\vct{r}_0$, Fig.~\ref{fig6} shows the characteristics of the estimated body displacement $\hat{d}_0(t)=(\lambda/4\pi)\angle I(t,\vct{r}_0)$; the dashed lines indicate abrupt body movements that were detected manually, where we observed that the target monkey was restless.

\begin{figure}[bt]
  \begin{center}
    \includegraphics[width=0.7\linewidth]{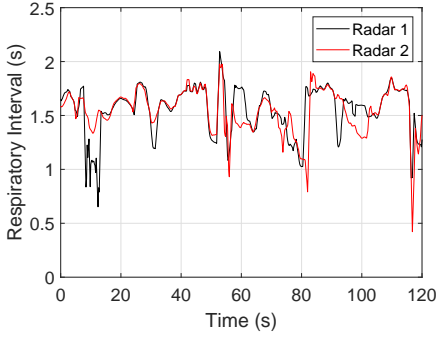}
    \caption{Respiratory intervals $\hat{\tau}_0(t)$ estimated using the conventional method from radar systems 1 (black) and 2 (red). The average error between the estimates from the two radar systems is 0.25 s.}
      \label{fig9}
  \end{center}
\end{figure}

Figure \ref{fig9} shows the respiratory intervals $\hat{\tau}_0(t)$ that were estimated from radar systems 1 (black) and 2 (red). In the figure, we see relatively large discrepancies between the estimates that are considered to be related to accuracy degradation caused by the body movements of the target monkey. Note that the respiratory interval $\hat{\tau}_0(t)$ was obtained from the second peak of $\rho_0(t,\tau)$ for $\vct{r}=\vct{r}_0$. The root mean square error between the respiratory intervals from radar systems 1 and 2 was as high as 0.25 s.

\begin{figure}[bt]
  \begin{center}
    \includegraphics[width=0.7\linewidth]{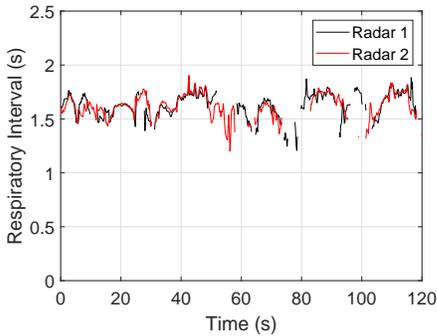}
    \caption{Respiratory intervals $\hat{\tau}(t)$ estimated using the proposed method for $\varepsilon_{\mathrm th}=0.5$ from radar systems 1 (black) and 2 (red). The average error between the estimates from the two radar systems is $0.08$ s.}
      \label{fig10}
  \end{center}
\end{figure}

Next, we apply the proposed method with a power threshold of $\eta=-20$ dB in ~(\ref{eq:4}), a window size of $T_0=2.0$ s in ~(\ref{eq:6}) and (\ref{eq:7}), and a residue threshold of $\varepsilon_{\mathrm th}=0.5$; these values were set empirically. The estimated respiratory intervals are presented in Fig.~\ref{fig10}. The root mean square (rms) error between the respiratory intervals from radar systems 1 and 2 was reduced to $0.08$ s, where the data acquisition rate was $84.4$\%. Note that the data acquisition rate $T_\mathrm{a}/T$ represents the ratio of the time $T_\mathrm{a}$ when the estimated respiratory interval was used to the total measurement time $T$.

Next, we reduced the threshold to $\varepsilon_{\mathrm th}=0.2$ and determined the respiratory intervals. The rms error between the respiratory intervals from radar systems 1 and 2 was reduced to $0.03$ s, indicating high accuracy. In contrast, the data acquisition rate was also reduced to $41.2$\%, which is the downside of use of the proposed method with a low threshold value.

\begin{figure}[tb]
  \begin{center}
    \noindent\begin{minipage}[b]{.47\linewidth} 
    \centering
    \hspace*{-5mm}\includegraphics[width=1.25\linewidth]{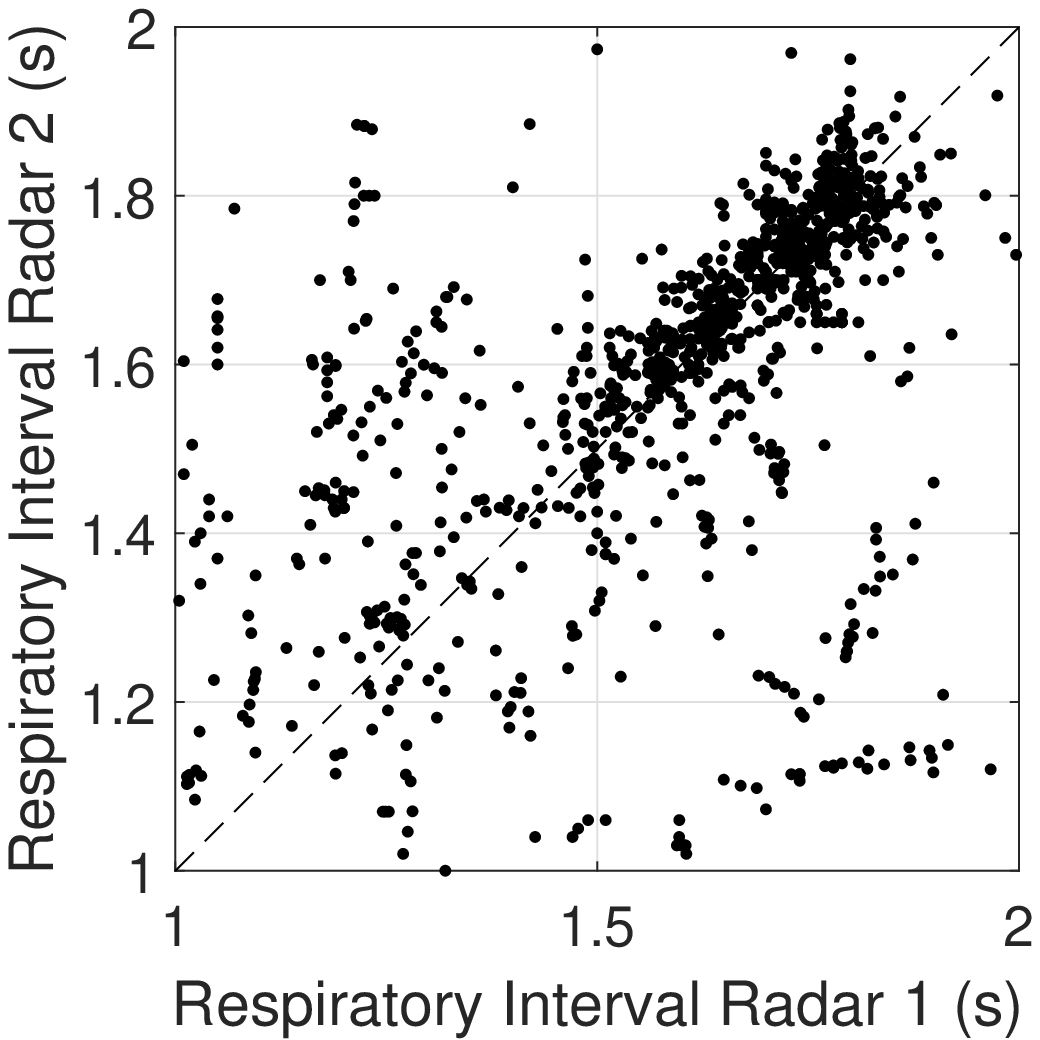}\\
    {(a) Conventional method ($C_\mathrm{cor}=0.55$).}
    \end{minipage}%
    \hspace*{0.01\linewidth}
    \begin{minipage}[b]{.47\linewidth}
      \centering
      \hspace*{-5mm}\includegraphics[width=1.25\linewidth]{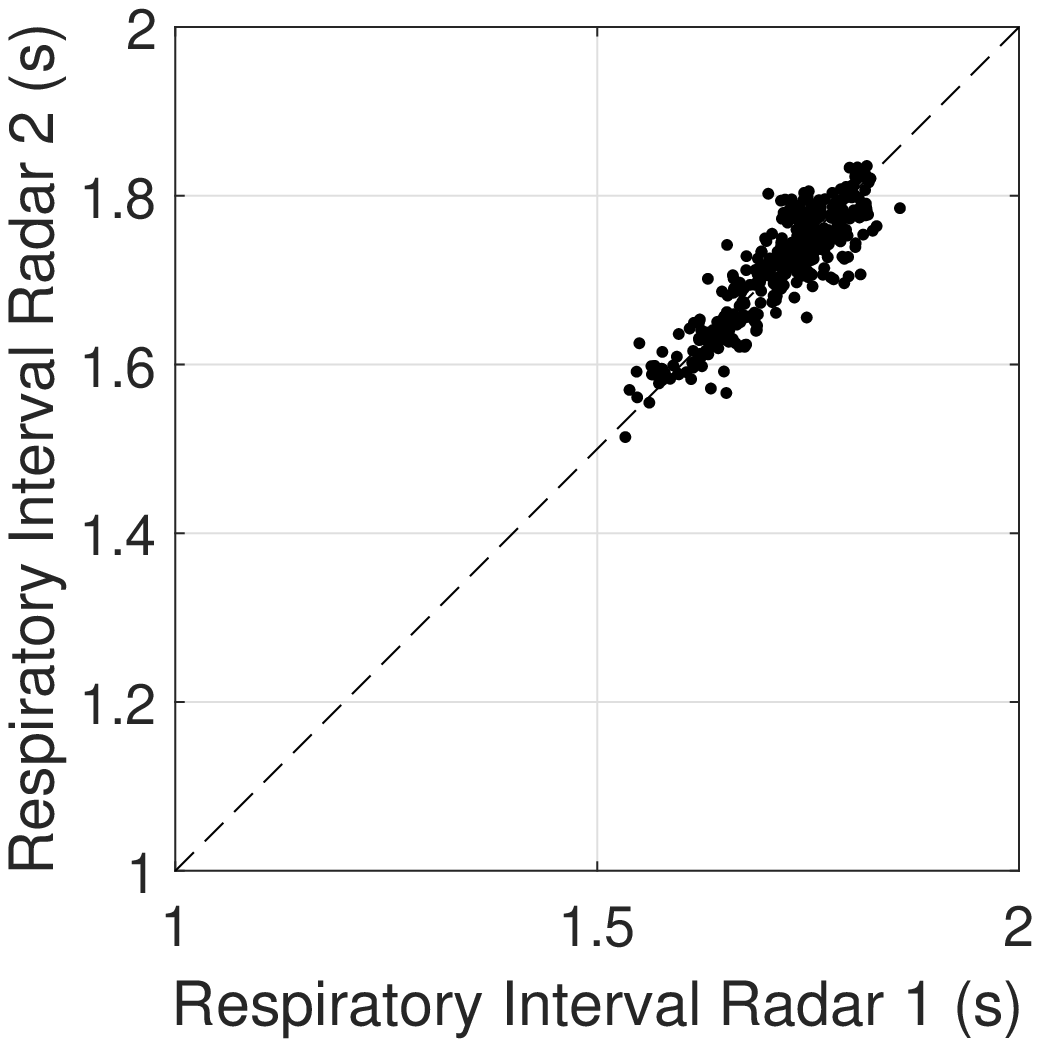}\\
     {(b) Proposed method ($C_\mathrm{cor}=0.89$).}
    \end{minipage}
    \caption{Scatter plots of $\hat{\tau}(t)$ for radar systems 1 and 2 when using (a) the conventional method and (b) the proposed method.}
     \label{fig12}
  \end{center}
\end{figure}

Fig. \ref{fig12} shows scatter plots of the respiratory intervals estimated using radar systems 1 and 2. Panels (a) and (b) correspond to $\tau_0(t)$ (from the conventional method) and $\tau(t)$ (from the proposed method with $\varepsilon_{\mathrm th}=0.2$), respectively. The correlation coefficients $C_\mathrm{cor}$ between the estimates obtained from radar systems 1 and 2 were $C_\mathrm{cor}=0.55$, $0.72$, and $0.89$ when using the conventional method, the proposed method with $\varepsilon_{\mathrm th}=0.5$, and the proposed method with $\varepsilon_{\mathrm th}=0.2$, respectively. This illustrates the effectiveness of the proposed method for accurate noncontact measurement of the monkey respiratory intervals using radar systems.

\section{Conclusion}
In this study, we have proposed a new method for radar-based noncontact measurement of the respiratory intervals of animals with frequent body movements. The proposed method uses a short-time correlation function of the echo phase to estimate the animal's respiratory interval. In addition, the similarity of the correlation function to a cosine function is used as a weight to average the respiratory interval estimates obtained from different distances and angles. After the averaging process is performed, the sum of weights is also used to select reliable estimates; based on this technique, we discard unreliable estimates of the respiratory intervals. As a result, the discrepancy between the respiratory intervals estimated using two radar systems decreased from $0.25$ s to $0.08$ s. In addition, the correlation coefficient between the estimates obtained from the two radar systems improved from $0.55$ to $0.89$. Our next important task will be to evaluate the accuracy of the proposed radar-based method when compared with contact-type sensors attached directly to the animal body; this will form part of our future studies.

\section*{Acknowledgment}
\addcontentsline{toc}{section}{Acknowledgment}
\scriptsize
This work was supported in part by SECOM Science and Technology Foundation, by JST under Grant JPMJMI22J2, and by JSPS KAKENHI under Grants 19H02155, 21H03427 and 23H01420. The authors thank all staff of Kyoto City Zoo who supported this study. The authors also thank Dr. Hirofumi Taki and Dr. Shigeaki Okumura of MaRI Co., Ltd. for their technical advice. This work involved animals in its research. Experiments on the animals in this study have been approved by the Research Ethics Committee of the Kyoto City Zoo (approval number: 2022-KCZ-006). 


\normalsize

\begin{thebibliography}{21}

\bibitem{JC_Lin1975}
  Lin J C (1975), ``Noninvasive microwave measurement of respiration,'' {\it Proc. IEEE,} vol. 63, pp. 1530--1530, doi: 10.1109/PROC.1975.9992.

\bibitem{S_Suzuki2009}
  Suzuki S, Matsui T, Kawahara H, Gotoh S (2009), ``Development of a noncontact and long-term respiration monitoring system using microwave radar for hibernating black bear,'' {\it Zoo Biol.,} vol. 28, pp.~259--270, doi: 10.1002/zoo.20229.

\bibitem{T_Matsumoto2022}
  Matsumoto T, Okumura S, Hirata S (2022), ``Non-contact respiratory measurement in a horse in standing position using millimeter-wave array radar,''
{\it J. Vet. Med. Sci.,} vol. 84, pp.~1340--1344.

\bibitem{SA_Tuan2022}
  Tuan S-A, Rustia D J A, Hsu J-T, Lin T-T (2022), ``Frequency modulated continuous wave radar-based system for monitoring dairy cow respiration rate,''
{\it Comput. Electron. Agric.,} vol. 196, 106913, doi: 10.1016/j.compag.2022.106913.

\bibitem{M_Nosrati2019}
  Nosrati M, Shahsavari S, Lee S, Wang H, Tavassolian N (2019), ``A concurrent dual-beam phased-array Doppler radar using MIMO beamforming techniques for short-range vital-signs monitoring,'' {\it IEEE Trans. Antennas Propag.}, vol. 67, pp. 2390--2404.

\bibitem{SMM_Islam2020}
  Islam S M M, Boric-Lubecke O, Lubekce V M (2020), ``Concurrent respiration monitoring of multiple subjects by phase-comparison monopulse radar using independent component analysis (ICA) with JADE algorithm and direction of arrival (DOA),'' {\it IEEE Access,} vol. 8, pp. 73558--73569, DOI: 10.1109/ACCESS.2020.2988038.
  
\bibitem{X_Shang2020} 
  Shang X, Liu J, Li J (2020), ``Multiple object localization and vital sign monitoring using IR-UWB MIMO radar,'' in {\it IEEE Trans. Aerosp. Electron. Syst.}, vol. 56, pp. 4437--4450.


\bibitem{J_Xiong2020} 
  Xiong J, Hong H, Zhang H, Wang N, Chu H, Zhu X (2020), ``Multitarget respiration detection with adaptive digital beamforming technique based on SIMO radar,'' {\it IEEE Trans. Microw. Theory Techn.}, vol. 68, pp. 4814--4824, DOI: 10.1109/TMTT.2020.3020082.
  
\bibitem{AA_Pramudita2021} 
Pramudita A A, Suratman F Y (2021), ``Low-power radar system for noncontact human respiration sensor,'' {\it IEEE Trans. Instrum. Meas.}, vol. 70, paper no. 4005415.

\bibitem{C_Feng2021} 
  Feng C {\it et al.} (2021), ``Multitarget vital signs measurement with chest motion imaging based on MIMO radar,'' {\em IEEE Trans. Microw. Theory Techn.,} vol. 69, pp. 4735--4747, doi: 10.1109/TMTT.2021.3076239.

\bibitem{M_Mercuri2021} 
  Mercuri M {\it et al.} (2021), ``Enabling robust radar-based localization and vital signs monitoring in multipath propagation environments,'' {\it IEEE Trans. Biomed. Eng.}, vol. 68, pp. 3228--3240.

\bibitem{Harikesh2021} 
  Harikesh, Chauhan S S, Basu A, Abegaonkar M P, Koul S K (2021), ``Through the wall human subject localization and respiration rate detection using multichannel Doppler radar,'' {\it IEEE Sensors J.,} vol. 21, pp. 1510--1518.
  
\bibitem{T_Zheng2022} 
Zheng T, Chen Z, Zhang S, Luo J (2022), ``Catch your breath: Simultaneous RF tracking and respiration monitoring with radar pairs,'' {\it IEEE Trans. Mobile Computing,} doi: 10.1109/TMC.2022.3197416.

\bibitem{A_Singh2021} 
  Singh A, Rehman S U, Yongchareon S, Chong P H J (2021), ``Multi-resident non-contact vital sign monitoring using radar: A review,'' {\it IEEE Sensors J.,} vol. 21, pp. 4061--4084.
  
\bibitem{C_Li2008} 
  Li C, Lin J (2008), ``Random body movement cancellation in Doppler radar vital sign detection,'' {\it IEEE Trans. Microw. Theory Techn.,} vol. 56, pp. 3143--3152, doi: 10.1109/TMTT.2008.2007139.
  
\bibitem{SH_Yu2020} 
  Yu S-H, Horng T-S (2020), ``Highly linear phase-canceling self-injection-locked ultrasonic radar for non-contact monitoring of respiration and heartbeat,'' {\it IEEE Trans. Biomed. Circuits Syst.}, vol. 14, pp. 75--90.
  
\bibitem{Q_Wu2021} 
  Wu Q, Mei Z, Lai Z, Li D, Zhao D (2021), ``A non-contact vital signs detection in a multi-channel 77 GHz LFMCW radar system,'' {\it IEEE Access}, vol. 9, pp. 49614--49628.
  
\bibitem{KC_Peng2021} 
  Peng K-C, Sung M-C, Wang F-K, Horng T-S (2021), ``Noncontact vital sign sensing under nonperiodic body movement using a novel frequency-locked-loop radar,'' {\it IEEE Trans. Microw. Theory Techn.}, vol. 69, pp. 4762--4773.

\bibitem{J_Zhang} 
  Zhang J, Yu J, Ma Y, Liang X (2021), ``RF-RES: Respiration monitoring with COTS RFID tags by Dopplershift,'' {\it IEEE Sensors J.}, vol. 21, pp. 24844--24854.
\end{thebibliography}
\end{document}